\documentclass[10pt,twocolumn]{article}

\usepackage[a4paper,margin=1.8cm]{geometry}
\usepackage[T1]{fontenc}
\usepackage{lmodern}
\usepackage{microtype}

\AtBeginDocument{%
  \DeclareFontShape{T1}{lmr}{m}{scit}{<->ssub*lmr/m/scsl}{}%
}

\usepackage{amsmath,amssymb}
\usepackage{graphicx}
\usepackage{eso-pic}
\usepackage{booktabs}
\usepackage{float}
\usepackage{stfloats}  
\usepackage{tikz}
\usetikzlibrary{arrows.meta, calc, positioning, shapes.geometric, fit, backgrounds, decorations.pathreplacing}
\usepackage{pgfplots}
\pgfplotsset{compat=1.18}
\usepackage[font=small]{caption}
\usepackage{enumitem}

\definecolor{oiBlue}{RGB}{0,114,178}
\definecolor{oiOrange}{RGB}{230,159,0}
\definecolor{oiGreen}{RGB}{0,158,115}
\definecolor{oiVermillion}{RGB}{213,94,0}

\definecolor{panelBg}{RGB}{250,250,250}
\definecolor{panelBdr}{RGB}{218,218,218}

\definecolor{hd}{RGB}{40,40,40}
\definecolor{bd}{RGB}{60,60,60}
\definecolor{dm}{RGB}{145,145,145}
\definecolor{ag}{RGB}{115,115,115}

\definecolor{rowG}{RGB}{237,250,244}
\definecolor{rowR}{RGB}{253,241,237}
\definecolor{pillG}{RGB}{225,245,235}
\definecolor{pillGb}{RGB}{0,158,115}
\definecolor{delRed}{RGB}{220,53,69}
\definecolor{delBg}{RGB}{253,237,237}

\usepackage{enumitem}
\setlist[itemize]{topsep=2pt,itemsep=1pt,parsep=0pt,partopsep=0pt,leftmargin=*}
\setlist[enumerate]{topsep=2pt,itemsep=1pt,parsep=0pt,partopsep=0pt,leftmargin=*}

\usepackage{titlesec}
\titlespacing*{\section}{0pt}{0.9ex plus 0.2ex minus 0.2ex}{0.5ex}
\titlespacing*{\subsection}{0pt}{0.7ex plus 0.2ex minus 0.2ex}{0.35ex}

\usepackage{algorithm}
\usepackage{algpseudocode}

\usepackage{amsthm}

\newtheorem{definition}{Definition}
\newtheorem{invariant}{Invariant}

\usepackage[hidelinks]{hyperref}
\usepackage{xurl}

\usepackage[numbers,sort&compress]{natbib}

\title{Securing Multi-Tool AI Agent Chains With Dynamic, Real-Time Compositional Policies}

\author{
  Chris Schneider$^{1}$ \quad Kriti Faujdar$^{2}$ \quad Philipp Schoenegger$^{1}$ \quad Ben Bariach$^{1}$ \\[1ex]
   \textnormal{$^{1}$Microsoft AI \quad\quad $^{2}$Microsoft}
}
\date{}

\begin{document}
 \twocolumn[
   \begin{@twocolumnfalse}
     \maketitle
   \end{@twocolumnfalse}
 ]

\AddToShipoutPictureBG*{%
  \AtPageUpperLeft{%
    \raisebox{-1.4cm}{\hspace{1.4cm}%
      \begin{minipage}{5cm}%
        \includegraphics[height=0.5cm]{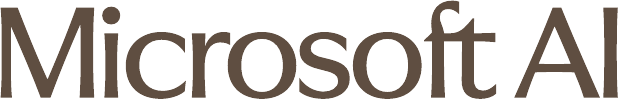}%
      \end{minipage}%
    }%
  }%
}

\begin{figure*}[!t]
  \centering
  \resizebox{\textwidth}{!}{%
  \begin{tikzpicture}

    \fill[blue!5, rounded corners=6pt] (-0.4,-4.6) rectangle (10.8,4.6);
    \node[font=\scriptsize\bfseries, blue!50!black, anchor=north west] at (-0.2,4.5) {PHASE 1: SESSION CHECKOUT};

    \fill[green!5, rounded corners=6pt] (11.6,-4.6) rectangle (23.4,4.6);
    \node[font=\scriptsize\bfseries, green!50!black, anchor=north west] at (11.8,4.5) {PHASE 2: RUNTIME EXECUTION};

    \draw[gray, line width=0.8pt, dashed] (11.2,-4.6) -- (11.2,4.6);


    \node[draw=oiBlue, fill=oiBlue!10, rounded corners=3pt, minimum width=7.0cm, minimum height=0.75cm, font=\small, align=center]
      (agent1) at (5.2,3.7) {\textbf{Agent} requests tools $T_1, \ldots, T_n$};

    \node[draw=oiBlue, fill=oiBlue!10, rounded corners=3pt, minimum width=7.0cm, minimum height=0.85cm, font=\small, align=center]
      (pstore) at (5.2,2.5) {\textbf{Policy Store}\\[-1pt]{\scriptsize\color{gray} per-tool: controls, data flow, network zones, TTL}};

    \draw[oiBlue!40, rounded corners=4pt, line width=0.5pt] (0.4,-1.9) rectangle (10.0,1.65);
    \node[font=\scriptsize\bfseries, oiBlue!70!black, anchor=north west] at (0.6,1.55) {MRS ENGINE};
    \node[font=\tiny, oiBlue!70!black, anchor=north east, fill=oiBlue!5, inner sep=2pt, rounded corners=1pt] at (9.8,1.55) {\texttt{mode}: Clearance $\mid$ Taint};

    \node[draw=oiBlue!70, fill=oiBlue!8, rounded corners=2pt, minimum width=8.0cm, minimum height=0.5cm, font=\scriptsize, align=center]
      (mrs1) at (5.2,0.95) {1.\ Compatibility check {\color{gray}\scriptsize (control conflicts, satisfiability)}};
    \node[draw=oiBlue!70, fill=oiBlue!8, rounded corners=2pt, minimum width=8.0cm, minimum height=0.5cm, font=\scriptsize, align=center]
      (mrs2) at (5.2,0.2) {2.\ Control resolution {\color{gray}\scriptsize $L_{\mathrm{eff}}(c)=\max_i(\mathrm{level}_i(c))$}};
    \node[draw=oiBlue!70, fill=oiBlue!8, rounded corners=2pt, minimum width=8.0cm, minimum height=0.5cm, font=\scriptsize, align=center]
      (mrs3) at (5.2,-0.55) {3.\ Data flow composition {\color{gray}\scriptsize (classification check, prohibition, zones)}};
    \node[draw=oiBlue!70, fill=oiBlue!8, rounded corners=2pt, minimum width=8.0cm, minimum height=0.5cm, font=\scriptsize, align=center]
      (mrs4) at (5.2,-1.3) {4.\ DENY enforcement {\color{gray}\scriptsize (composed flow vs.\ deny-level controls)}};

    \node[draw=oiBlue!60!black, fill=white, rounded corners=2pt, minimum width=6.0cm, minimum height=0.5cm, font=\scriptsize, align=center]
      (ecs) at (5.2,-2.35) {$\to$ \textbf{Effective Control Set}};

    \node[draw=oiGreen, fill=oiGreen!12, rounded corners=3pt, minimum width=3.0cm, minimum height=0.65cm, font=\small, align=center]
      (session) at (2.8,-3.8) {\textbf{Session} created};
    \node[draw=oiVermillion, fill=oiVermillion!12, rounded corners=3pt, minimum width=3.0cm, minimum height=0.65cm, font=\small, align=center]
      (blocked) at (7.6,-3.8) {\textbf{Blocked}};

    \draw[-{Stealth[length=5pt]}, oiBlue!70!black, line width=0.6pt] (agent1) -- (pstore);
    \draw[-{Stealth[length=5pt]}, oiBlue!70!black, line width=0.6pt] (pstore) -- (mrs1);
    \draw[-{Stealth[length=5pt]}, oiBlue!70!black, line width=0.6pt] (mrs1) -- (mrs2);
    \draw[-{Stealth[length=5pt]}, oiBlue!70!black, line width=0.6pt] (mrs2) -- (mrs3);
    \draw[-{Stealth[length=5pt]}, oiBlue!70!black, line width=0.6pt] (mrs3) -- (mrs4);
    \draw[-{Stealth[length=5pt]}, oiBlue!70!black, line width=0.6pt] (mrs4) -- (ecs);
    \draw[-{Stealth[length=5pt]}, oiGreen!70!black, line width=0.6pt] (ecs.south) -- ++(0,-0.35) -| (session.north);
    \draw[-{Stealth[length=5pt]}, oiVermillion, line width=0.6pt]     (ecs.south) -- ++(0,-0.35) -| (blocked.north);
    \node[font=\scriptsize, oiGreen!70!black] at (3.6,-3.2) {valid};
    \node[font=\scriptsize, oiVermillion]     at (6.8,-3.2) {invalid};


    \node[draw=oiGreen!70!black, fill=oiGreen!8, rounded corners=3pt, minimum width=7.0cm, minimum height=0.75cm, font=\small, align=center]
      (agent2) at (17.2,3.7) {\textbf{Agent} calls tool $T_i$ on resource $r$};

    \node[draw=oiGreen!70!black, fill=oiGreen!8, rounded corners=3pt, minimum width=7.0cm, minimum height=0.85cm, font=\small, align=center]
      (guard) at (17.2,2.5) {\textbf{Session Guard}\\[-1pt]{\scriptsize\color{gray} active, not expired, tool in checked-out set}};

    \node[draw=oiGreen!70!black, fill=oiGreen!8, rounded corners=3pt, minimum width=7.0cm, minimum height=0.85cm, font=\small, align=center]
      (classify) at (17.2,1.2) {\textbf{Resource Classifier}\\[-1pt]{\scriptsize\color{gray} look up $r$ in label catalog $\to$ classification label}};

    \node[draw=oiGreen!70!black, fill=oiGreen!8, rounded corners=3pt, minimum width=7.0cm, minimum height=0.85cm, font=\small, align=center]
      (taint) at (17.2,-0.1) {\textbf{Taint Tracker}\\[-1pt]{\scriptsize\color{gray} update session high-water mark (sticky prohibit-external)}};

    \node[draw=oiGreen!60!black, fill=white, rounded corners=2pt, minimum width=5.0cm, minimum height=0.5cm, font=\scriptsize, align=center]
      (check) at (17.2,-1.3) {enforcement checks (\S\ref{sec:runtime-model})};

    \node[draw=oiGreen, fill=oiGreen!12, rounded corners=3pt, minimum width=3.0cm, minimum height=0.65cm, font=\small, align=center]
      (execute) at (14.8,-3.8) {\textbf{Tool executes}};
    \node[draw=oiVermillion, fill=oiVermillion!12, rounded corners=3pt, minimum width=3.0cm, minimum height=0.65cm, font=\small, align=center]
      (killed) at (19.6,-3.8) {\textbf{Session revoked}};
 
    \draw[-{Stealth[length=5pt]}, oiGreen!70!black, line width=0.6pt] (agent2) -- (guard);
    \draw[-{Stealth[length=5pt]}, oiGreen!70!black, line width=0.6pt]             (guard) -- (classify);
    \draw[-{Stealth[length=5pt]}, oiGreen!70!black, line width=0.6pt] (classify) -- (taint);
    \draw[-{Stealth[length=5pt]}, oiGreen!70!black, line width=0.6pt] (taint) -- (check);
    \draw[-{Stealth[length=5pt]}, oiGreen!70!black, line width=0.6pt] (check.south) -- ++(0,-0.8) -| (execute.north);
    \draw[-{Stealth[length=5pt]}, oiVermillion, line width=0.6pt]     (check.south) -- ++(0,-0.8) -| (killed.north);
    \node[font=\scriptsize, oiGreen!70!black] at (15.6,-2.55) {no};
    \node[font=\scriptsize, oiVermillion]     at (18.8,-2.55) {yes};

    \draw[-{Stealth[length=5pt]}, gray, line width=0.6pt, rounded corners=5pt]
      (execute.south) -- (14.8,-4.3) -- (22.8,-4.3) -- (22.8,3.7) -- (agent2.east);
    \node[font=\scriptsize, gray, rotate=90, anchor=south] at (23.2,0.0) {next tool call};

  \end{tikzpicture}%
  }
  \caption{DSCC architecture overview. Phase~1 (left) uses the MRS composition engine, operating in one of two modes (\emph{Clearance} or \emph{Taint}), to compose per-tool policies into a single Effective Control Set at session checkout.}
  \label{fig:architecture}
\end{figure*}
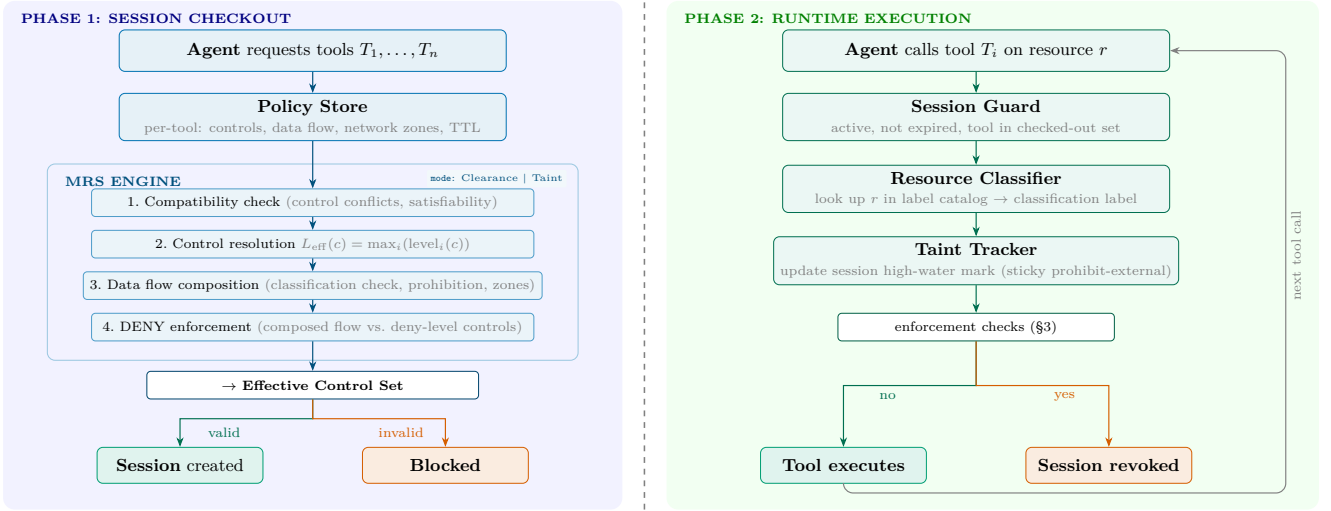


\begin{abstract}
\textit{%
Modern AI agent implementations such as frontier coding agents chain multiple tools at runtime that create a security surface that per-tool guardrails are unable to address, as individually permitted tools can violate organizational policies when composed. We propose the Dynamic Security Control Compositor (DSCC), a two-phase approach to compositional security for multi-tool agent chains. In Phase~1, at session checkout, a Most Restrictive Set (MRS) algorithm composes per-tool security policies into a single effective policy for the full chain with a formal monotonicity invariant that extending a chain can only tighten the result, blocking incompatible combinations before any tool executes. Outputs of any tool call propagate their classification constraints into a session-level taint state, so subsequent invocations must satisfy the most restrictive classification constraints seen so far on the chain. In Phase~2, at runtime, the system tracks the sensitivity of data the agent touches through a monotonic taint state and revokes the session if the accumulated exposure would make a subsequent tool call a policy violation. Together, these two phases provide defense in depth, where static composition prevents unsafe chains from starting, and runtime taint tracking catches violations that emerge from the specific data used. We then provide a reference implementation on a catalog of 32 tools governed by 16 NIST SP~800-53 aligned policies and evaluate it under two composition modes. In the default \emph{clearance} mode, permitted combinations are partitioned into classification-level clusters, blocking 79.2\% of policy pairs and 95.5\% of triples. The alternative \emph{taint} mode admits mixed-classification chains within the exfiltration boundary, blocking 42.5\% and 60.5\% respectively. Lastly, we discuss the governance implications for organizations deploying multi-tool agents, including the utility-security tradeoff and the changes needed to operationalize chain-aware policies.
}
\end{abstract}


\section{Introduction}

Large language model agents can invoke external tools, ranging from reading files and calling APIs to sending messages and executing code.
This has been shown by Schick et al.~\cite{schick2023toolformer} to be able to be done autonomously with models as early as GPT-J. Later, Yao et al.~\cite{yao2023react} then demonstrated that interleaving reasoning with tool actions substantially improves metrics such as hallucination rates, error propagation, and interpretable output, leading to improved performance and applicability. This has contributed to the widespread application of LLMs across a large set of knowledge work due to the added capabilities and reduced error rates. Subsequent work then scaled tool use to thousands of real-world APIs~\cite{patil2023gorilla, qin2023toolllm}, leading to modern deployed agentic coding tools such as GitHub Copilot~\cite{github2025copilotagent}, Claude Code~\cite{anthropic2025claudecode}, and Codex~\cite{openai2025codex}, which give LLMs direct access to a substantial number of file systems, shells, and web APIs, making multi-tool composition a standard mode of operation in pursuit of the user goals, which often include a wide range of sometimes low-level computer operations. Moreover, as agent teams, cross-ecosystem invocations, and agent-swarms become more common-place and capabilities increase even further, the total number of tools and tool chains used in each session is likely to increase dramatically. 

We argue that this creates a security surface that current safeguards are not well designed to address and that is likely to become more pressing in the near future. Consider a simple example: a file-reading tool and an HTTP client each pass their individual security checks, but when composed, an agent can read an internal document and POST its contents to an external server. Neither tool is prohibited on its own, with the violation only emerging from their combination. Nor is the risk limited to data exfiltration, with a database query tool handling Confidential data (requiring AC-6 Least Privilege) and a planning tool classified as Internal (carrying only AC-3 Access Enforcement) each operating correctly in isolation. However, composing them means the planning tool would process Confidential data without the access controls mandated at that classification level, even though no data ever leaves the organization.
The context is the component that necessitates security policy alignment versus the tool. Scaled up to multiple sub-agents and large agent teams across multiple sessions, the combinations of tools are likely to result in severe security risks. 

Recent work shows that this type of risk is already materializing. For example, Li et al.~\cite{li2025stac} formalize what they call Sequential Tool Attack Chaining (STAC), constructing multi-turn attack sequences in which each individual attack prompt has less than 2\% harmfulness, yet the composed sequence achieves an attack success rate exceeding 90\% across eight LLMs spanning four model families, with their strongest defense reducing this by only 28.8\%.
Moreover, Wu et al.~\cite{wu2026chainfuzzer} apply greybox fuzzing to 20 open-source agent applications covering 998 tools, finding that 302 of 365 confirmed vulnerabilities (82.7\%) require multi-tool execution to manifest, with only 63 being single-tool issues, resulting in an almost 5 times higher incidence of multi-tool vulnerabilities. This suggests that multi-tool use is a frequent cause of vulnerabilities in current implementations, further strengthening the central security problem of these combinatorial risks. 

A growing body of work on the Model Context Protocol (MCP) also corroborates this pattern, documenting cross-server memory exfiltration~\cite{sun2025msa}, data leakage via implicit prompt injection that evades output-based safety checks in 95\% of cases~\cite{lan2026silentegress}, implicit tool poisoning that manipulates agents into misusing legitimate high-privilege tools~\cite{li2026mcpitp}, tool poisoning as the most prevalent client-side vulnerability under STRIDE/DREAD analysis of Huang et al. ~\cite{huang2026mcpthreat}, and over 2{,}000 attack instances across 10 scenarios and 12 attack categories~\cite{zhang2025mcpsecuritybench}. As agents often autonomously interact with MCPs directly, this directly interlinks with the above risk surface.

While much of this recent work targets MCP specifically, the underlying vulnerability is protocol-agnostic, with any system in which an agent sequences tool calls, whether through OpenAI function calling~\cite{openai2023functioncalling}, LangChain~\cite{chase2022langchain}, AutoGen~\cite{wu2023autogen}, or bespoke orchestration, is facing the same compositional risk whenever per-tool policies are evaluated in isolation. In general, stochastic software such as LLM agents can invoke arbitrary tools that require per-invocation authorization, which determines whether a single call satisfies the policy active at that specific point. Additionally, they also require chain-level composition, which intersects the policies of every tool traversed by the agent and rejects any path that has an effective permission which exceeds that intersection. Many existing frameworks implement the first and largely omit the second, which is exactly the compositional gap this paper addresses.

In this paper, we propose a solution to the compositional authorization gap in agent tool-chains by making four distinct contributions. To remediate these issues, we introduce the Dynamic Security Control Compositor (DSCC), a two-phase architecture for compositional security of multi-tool agent chains (Figure~\ref{fig:architecture}). DSCC treats the agent as an untrusted stochastic caller whose tool selections cannot be predicted at session time, making the threat not a specific adversary but the emergent policy violations that arise when individually-authorized tool calls compose into sequences whose aggregate effective permissions exceed what any constituent policy intended to grant.

Our approach includes the requirement that every tool carry a standardized security policy encoding its data classification, permitted data flows, as well as applicable regulatory controls, resulting in the common representation required for composition. We then define the Most Restrictive Set (MRS) algorithm as the composition operator at the core of DSCC's Phase~1, which takes the per-tool policies of any chain and produces a single effective policy, which satisfies a monotonicity invariant such that extending a chain can only tighten the result but never relax it. We took the basic idea of ordered security labels from Bell–LaPadula's security model~\cite{bell1973blp} and adapted it for agent chains by treating agents as control subjects, enforcing monotonic tightening of policies as the chain grows and introducing a compositional algebra for multi-tool policy merging (novel to this framework). Our approach also removes trusted-subject exemptions.

Specifically, our proposed Phase~1 applies this composition at session checkout, where, in its default clearance mode, it rejects incompatible chains before they execute. In its other mode, the alternative taint mode, it permits mixed-classification chains while propagating the chain's high-water mark forward as session state. Phase~2 then extends the same taint state at runtime, where it revokes any given session session before any subsequent call would breach it. 

Together, this static composition prevents unsafe chains from starting, while runtime taint tracking catches violations that emerge from the specific data encountered during an agent's runtime. We evaluate this approach through a reference .NET~9 implementation over 32~tools governed by 16~NIST SP~800-53 aligned policies. We find that under the clearance mode, a total of 79.2\%/95.5\% of policy pairs/triples are blocked, while under the taint mode it is 42.5\%/60.5\%, in line with our theoretical expectations. Lastly, we discuss the practical governance implications of compositional security for multi-tool agent deployments and what organizational changes are needed to operationalize it.


\section{The MRS Algorithm}
\label{sec:mrs}

This section presents the Most Restrictive Set (MRS) algorithm, which powers Phase~1 of the DSCC architecture. First, we define the policy model each tool carries (\S\ref{sec:policy-model}). Second, we present the four-step composition procedure that produces a single effective control set for the chain (\S\ref{sec:composition}). Third, we state the monotonicity invariant guaranteeing that adding a tool can never relax the result (\S\ref{sec:monotonicity}). Jointly, these components ensure that any tool chain is either governed by a single policy at least as restrictive as each individual tool's, or rejected with a clear explanation before execution begins.


\subsection{Policy Model}
\label{sec:policy-model}

In order for the Most Restrictive Set (MRS) algorithm to function, it requires a common representation of what each tool is permitted to do in the same format. This allows the constraints from different tools to be compared and combined mechanically in a straightforward manner. Most agent frameworks today define tools primarily by their functional capability (e.g., 'a web search tool'), with limited or no security metadata attached. The proposed DSCC solution directly shapes this layer, such that every tool registered in the system carries a security policy authored by the tool developer or organizational security team, encoding what data the tool may touch, where it may send that data, and which NIST SP~800-53 controls apply to its operation. Definition~\ref{def:tool-policy} specifies the six components of this policy.

\begin{definition}[Tool Policy]
\label{def:tool-policy}
A \emph{tool policy} $P_i$ for tool $T_i$ is the six-tuple
\[
  P_i = \bigl(\mathcal{B}_i,\; C_i,\; \phi_i,\; d_i,\; Z_i,\; \tau_i\bigr).
\]

\smallskip
\small
\captionsetup{hypcap=false}%
\captionof{table}{Components of the tool-policy six-tuple $P_i$ (Definition~\ref{def:tool-policy}).}%
\label{tab:policy-components}%
\begin{tabular}{@{}c p{2.3cm} p{3.8cm}@{}}  
\toprule
\textbf{Symbol} & \textbf{Component} & \textbf{Definition} \\
\midrule
$\mathcal{B}_i$ & Control bindings &
  Pairs $(c,l)$ binding a NIST control $c$ (e.g.\ AC-4) to $l \in \{\textsc{Allow},\textsc{Restrict},\textsc{Deny}\}$. \\[3pt]
$C_i$ & Data-flow classification &
  Highest sensitivity handled (from \textsc{Public} up to \textsc{Restricted}). \\[3pt]
$\phi_i$ & Transmission prohibition &
  Flag in $\{0,1\}$; if $1$, $T_i$'s data must stay within the boundary. \\[3pt]
$d_i$ & Flow direction &
  Sends out, receives, both, or internal-only. \\[3pt]
$Z_i$ & Permitted zones &
  Zones $T_i$ may run in (\texttt{corporate-vpn}, \texttt{dmz}, \dots); any if unset. \\[3pt]
$\tau_i$ & Maximum session TTL &
  Longest session lifetime, in hours (default 48). \\
\bottomrule
\end{tabular}
\end{definition}

The first three components capture what the tool does with data. First, control bindings tie NIST controls to restriction levels that determine whether an operation is permitted, audited, or blocked. Second, the classification level records the highest sensitivity of data the tool handles. Third, the prohibition flag marks whether that data must stay within the organizational boundary. The restriction and classification levels are both totally ordered, which lets the composition algorithm select the most restrictive value mechanically.

The remaining three components capture where and how long the tool operates. First, the flow direction declaring the tool's data movement capability using combinable flags. Second, the permitted zones constraining which network segments it may run in. Third, (vi) the TTL setting an upper bound on session lifetime.

\begin{table*}[!t]
\centering
\caption{Domains used in tool policies (Definition~\ref{def:tool-policy}).}
\small
\begin{tabular}{@{}lp{9.5cm}l@{}}
\toprule
\textbf{Domain} & \textbf{Ordering / Semantics} & \textbf{Interpretation} \\
\midrule
Restriction level & $\textsc{Allow} < \textsc{Restrict} < \textsc{Deny}$ & Permissive $\to$ prohibitive \\[3pt]
Classification level & $\textsc{Public} < \textsc{Internal} < \textsc{Confidential} < \textsc{Restricted}$ & Low $\to$ high sensitivity \\[3pt]
Flow direction & Flags: \textsc{Inbound}, \textsc{Outbound}, \textsc{Bidirectional}, \textsc{InternalOnly} & Data movement capability \\
\bottomrule
\end{tabular}
\smallskip

{\footnotesize \textit{Note.} Restriction levels and classification levels are ordered from least to most restrictive such that any two values can be compared, while flow directions are combinable flags describing data movement capability, not an ordered scale.}
\end{table*}

The policy model requires that, first, a tool $T_i$ is \emph{outbound} if its flow direction includes the outbound flag but not the internal-only flag (as the composition algorithm uses this predicate to identify tools capable of sending data outside the organizational boundary) and ,second, controls not mentioned in a tool's policy are simply absent from that tool's binding set, as they do not default to \textsc{Allow} or \textsc{Deny} and do not participate in control resolution for that tool.


\subsection{Composition Algorithm}
\label{sec:composition}

This section describes how policies from multiple tools are composed into a single effective control set. Given a chain of tools $T_1, \ldots, T_n$ with policies $P_1, \ldots, P_n$, the MRS algorithm produces either an \emph{effective control set} $E$ that governs the entire session or a rejection with an attributed reason that clearly identifies which policies conflict and why. This approach proceeds in four steps, each of which can independently reject the chain: compatibility check, control resolution, data-flow composition, and DENY enforcement. 

\begin{table*}[t]
\caption{MRS Composition}
\label{alg:mrs}
\centering
\begin{minipage}{0.85\textwidth}
\hrule
\vspace{5pt}
\begin{algorithmic}[1]
\Require Policies $P_1, \ldots, P_n$ for tools $T_1, \ldots, T_n$;
         composition mode $mode \in \{\textsc{Clearance},\, \textsc{Taint}\}$ (default: \textsc{Clearance});
         optional initial classification $C_0$
\Ensure Effective control set $E$ or \textsc{Reject}
\Statex
\State \textbf{Step 1:} Check structural compatibility of policy set
\Statex
\State \textbf{Step 2:} For each control $c$ declared by any tool: $L_{\mathrm{eff}}(c) \gets \max_i \, l_i(c)$
\Statex
\State \textbf{Step 3:} Compose data flow:
\State \quad $C_{\mathrm{eff}} \gets \max\bigl(C_0,\; \max_i \, C_i\bigr)$ \Comment{high-water mark}
\State \quad \textbf{if} $mode = \textsc{Clearance}$: Reject if any $C_i < C_{\mathrm{eff}}$ \Comment{clearance check}
\State \quad $\phi_{\mathrm{eff}} \gets \max_i \, \phi_i$ \Comment{sticky prohibition}
\State \quad Reject if $\phi_{\mathrm{eff}} = 1$ and chain contains outbound tools
\State \quad Reject if $C_{\mathrm{eff}} \geq \textsc{Confidential}$ and chain contains outbound tools
\State \quad $Z_{\mathrm{eff}} \gets \bigcap_i Z_i$ \Comment{zone intersection}
\State \quad Reject if $Z_{\mathrm{eff}} = \emptyset$
\Statex
\State \textbf{Step 4:} For each boundary or flow control $c$ at \textsc{Deny}: reject if composed flow violates $c$
\Statex
\State $\tau_{\mathrm{eff}} \gets \min_i \, \tau_i$
\State \Return $E = \bigl(\{(c, L_{\mathrm{eff}}(c))\},\; C_{\mathrm{eff}},\; \phi_{\mathrm{eff}},\; Z_{\mathrm{eff}},\; \tau_{\mathrm{eff}}\bigr)$
\end{algorithmic}
\vspace{5pt}
\hrule
\vspace{6pt}
{\footnotesize
\textbf{Notation.}\;
$T_i$: a tool in the chain.\;
$P_i$: the security policy attached to $T_i$ (Definition~\ref{def:tool-policy}).\;
$c$: a NIST SP~800-53 control identifier (e.g., AC-4, SC-7).\;
$l_i(c)$: the restriction level (\textsc{Allow} $<$ \textsc{Restrict} $<$ \textsc{Deny}) that $T_i$'s policy assigns to control $c$.\;
$L_{\mathrm{eff}}(c)$: the effective (most restrictive) level of control $c$ across the chain.\;
$C_i$: data-flow classification (\textsc{Public} $<$ \textsc{Internal} $<$ \textsc{Confidential} $<$ \textsc{Restricted}).\;
$\phi_i$: transmission prohibition flag (1\,=\,data must stay internal).\;
$Z_i$: set of permitted network zones; $\bigcap_i Z_i$ is their intersection (zones common to all tools).\;
$\tau_i$: maximum session lifetime.\;
$E$: the resulting effective control set for the chain.\;
$C_0$: optional initial classification of the input data (default: \textsc{Public}); seeds the high-water mark before any tool contributes.\;
$mode$: composition mode (\textsc{Clearance} or \textsc{Taint}); governs whether the clearance check in Step~3 is enforced.\;
Subscript $_{\mathrm{eff}}$ denotes the composed value for the full chain.
}
\end{minipage}
\end{table*}

\paragraph{Step 1: Compatibility check.}

Before computing anything, the first step of this algorithm verifies that the requested tool set is not internally contradictory. Some policy combinations are unsatisfiable by themselves, regardless of composition order. For example, if one policy prohibits external transmission ($\phi_i = 1$) while another explicitly permits operation on the public internet, the two are contradictory, as one tool requires data to stay internal while another sends it externally. Step~1 also delegates an early data-flow satisfiability check to the composer (Step~3). If either check fails, the chain is rejected before composition begins. \footnote{In deployments, chains can become too long such that pathological compositions become common. In these cases, operators may want to decompose the workflow into smaller sub-chains, each independently composed and checked out. This narrows the effective control set per segment and aligns with the modular-design principle of small, auditable units.}

\paragraph{Step 2: Control resolution.}

Next, for each NIST control $c$ appearing in any policy in the chain, the algorithm selects the most restrictive properties across all tools that declare it. Writing $l_i(c)$ for the restriction properties that tool $T_i$'s policy assigns to control $c$:
\begin{equation}
\label{eq:control-resolution}
  L_{\mathrm{eff}}(c) \;=\; \max_{\substack{i \,:\, T_i \text{ declares } c}} l_i(c)
\end{equation}
A single \textsc{Deny} from any tool in the chain overrides, as the chain as a whole must satisfy the strictest individual requirement. This contrasts with a majority vote that would allow permissive tools to dilute a restrictive one. The resolved control record then also stores which policy imposed the winning level, providing attribution for audit and incident response.

\paragraph{Step 3: Data-flow composition.}

This step composes the data-flow constraints from all tools in the chain, computing five properties that together determine how data may flow through the session and whether that flow is permitted.

\emph{Classification high-water mark.}\quad The effective classification for the session is set by whichever tool handles the most sensitive data, taking the maximum classification level across all tools in the chain:
\begin{equation}
\label{eq:classification}
  C_{\mathrm{eff}} \;=\; \max\!\bigl(C_0,\; \max_i \, C_i\bigr)
\end{equation}
This value can only increase as tools are added to the chain. For example, if one tool handles \textsc{Confidential} data, the entire session must be governed at that classification level, because the agent may route that tool's output to any subsequent tool. Equivalently, classification sticks with the session through the chain. We compute this up front as the high-water mark, and \S\ref{sec:runtime-model} shows how the session taint state updates it as the chain runs. The high-water mark is seeded from $C_0$ with the classification of the input data if known or, if not available, starts at \textsc{Public}.

\emph{Clearance check.}\quad MRS supports two composition modes that differ only in whether the clearance check is enforced. In \emph{clearance mode} (the default), tools require a classification level at least as high as $C_{\mathrm{eff}}$, otherwise the chain is rejected because those tools are not cleared to handle data at the session's sensitivity level. For example, a \textsc{Public}-classified HTTP client cannot participate in a chain that touches \textsc{Confidential} data.

In \emph{taint mode}, the optional setting, the clearance check is skipped. Classification is treated as a sticky property of the data rather than a property the tool must be pre-authorized for. For example, a\textsc{Public}-classified HTTP client may process \textsc{Confidential} data, provided the output remains classified as \textsc{Confidential} (the upward ratchet still applies). All other checks—prohibition propagation, the classification boundary rule, and zone intersection—remain in force irrespective of which mode is chosen. In general, we argue that the taint mode is suited to mixed-classification pipelines where individual tools are not independently credentialed for the chain's peak sensitivity, but organizational controls do govern the overal pipeline.

\emph{Prohibition propagation.}\quad Once any tool in the chain prohibits external transmission, that prohibition binds the entire session and cannot be lifted by subsequent tools:
\begin{equation}
\label{eq:prohibition}
  \phi_{\mathrm{eff}} \;=\; \max_i \, \phi_i
\end{equation}
For example, if a file reader prohibits external transmission, that prohibition propagates to the entire session, blocking any subsequent outbound tool from sending the data externally.

\emph{Classification boundary rule.}\quad Even without an explicit prohibition flag, any chain carrying data classified at \textsc{Confidential} or above that also contains an external-facing tool is rejected. This provides a safety floor, ensuring that sensitive data must not reach external tools even if no individual policy explicitly sets $\phi_i = 1$.

\emph{Network zone intersection.}\quad Each tool declares which network zones it may operate in, and the session is restricted to zones that every tool permits:
\begin{equation}
\label{eq:zones}
  Z_{\mathrm{eff}} \;=\; \bigcap_i Z_i
\end{equation}
where a null $Z_i$ (unrestricted) is treated as the universal set. If $Z_{\mathrm{eff}} = \emptyset$, no network zone satisfies all tools simultaneously and the chain is rejected.

Zone intersection is also the one composition operator that can produce failures invisible to pairwise analysis. For instance, consider three tools with $Z_1 = \{A, B\}$, $Z_2 = \{B, C\}$, $Z_3 = \{A, C\}$, where every pair shares a zone but $Z_1 \cap Z_2 \cap Z_3 = \emptyset$, so the triple is rejected even though all pairs pass. Because of this, our architecture composes over the full tool set rather than checking pairs.\footnote{For the other operators (maximum, minimum, Boolean OR), the pairwise outcome already determines the $n$-way outcome because the ``non-violating'' predicate is preserved under repeated application. For zone intersection, all operators are equally associative, but pairwise non-emptiness does not imply $n$-way non-emptiness, which is what makes it the structural source of pairwise-invisible failures.}

If any of the checks in this step fails, the composer returns a rejection identifying the conflicting tools and the NIST control that was violated.

\paragraph{Step 4: DENY enforcement.}

Steps~2 and~3 can each independently reject a chain, but some violations require information from both. During the last step, Step~4 cross-checks the resolved controls against the composed data flow. For each control at \textsc{Deny} level in the resolved set, the algorithm checks whether the composed data flow would violate it. Specifically, if the chain contains external-facing tools and the effective classification exceeds \textsc{Public}, any control at \textsc{Deny} level that governs data boundary crossing or information flow triggers rejection.

Note that a chain that survived Steps~1--3 can still fail here. For instance, two tools might have a boundary control at \textsc{Restrict}, but a third tool escalates it to \textsc{Deny} via Step~2. Only at Step~4, when the DENY-level control is checked against the composed data flow, does the violation surface.

If all four steps succeed, the effective session lifetime is computed as $\tau_{\mathrm{eff}} = \min_i \, \tau_i$, inheriting the shortest TTL of any constituent tool. The resulting effective control set $E$ then governs the session for its duration. Table~\ref{alg:mrs} summarizes the full computation.


\subsection{Monotonicity Invariant}
\label{sec:monotonicity}

The composition operators defined in \S\ref{sec:composition} share a structural property, which is that because each uses maximum, minimum, or intersection, extending a chain can never relax the effective control set. We state this as a formal invariant below.

\begin{invariant}[Monotonicity]
\label{inv:monotonicity}
Let $E_n$ be the effective control set for a chain of $n$ tools
$\langle T_1, \ldots, T_n \rangle$, and let $E_{n+1}$ be the effective control
set after appending tool $T_{n+1}$. Then $E_{n+1}$ is at least as restrictive as
$E_n$ on every dimension:

\smallskip
\noindent\begin{minipage}{\linewidth}
\captionsetup{hypcap=false}%
\captionof{table}{Monotonicity of the effective control set on each policy dimension under chain extension (Invariant~\ref{inv:monotonicity}).}%
\label{tab:monotonicity}%
{\normalfont\small
\begin{tabular}{@{}l c l@{}}
\toprule
\textbf{Dimension} & \textbf{Dir.} & \textbf{Relation ($n \to n{+}1$)} \\
\midrule
Restriction level $L_{\mathrm{eff}}$ & $\uparrow$   & $L_{\mathrm{eff}}^{(n+1)}(c) \geq L_{\mathrm{eff}}^{(n)}(c)$ \\[3pt]
Classification $C_{\mathrm{eff}}$    & $\uparrow$   & $C_{\mathrm{eff}}^{(n+1)} \geq C_{\mathrm{eff}}^{(n)}$ \\[3pt]
Network zones $Z_{\mathrm{eff}}$     & $\downarrow$ & $Z_{\mathrm{eff}}^{(n+1)} \subseteq Z_{\mathrm{eff}}^{(n)}$ \\[3pt]
Session TTL $\tau_{\mathrm{eff}}$    & $\downarrow$ & $\tau_{\mathrm{eff}}^{(n+1)} \leq \tau_{\mathrm{eff}}^{(n)}$ \\[3pt]
Prohibition $\phi_{\mathrm{eff}}$    & $\uparrow$   & $\phi_{\mathrm{eff}}^{(n+1)} \geq \phi_{\mathrm{eff}}^{(n)}$ \\
\bottomrule
\end{tabular}}

\smallskip
{\footnotesize\textit{Note.} In this table, the arrows up/down indicate the direction each quantity may move. The superscript $(n)$ denotes the chain of length~$n$.}
\end{minipage}
\end{invariant}

This means that based on this, organizations can make decisions about tool additions in a modular manner. That is, if a chain is currently approved, and because adding a tool can only either preserve of tighten the effective control set), the decision-making is focused on the additional step as opposed to the full chain. As such, any reviewer need only assess whether the new tool's contribution is acceptable rather than re-auditing the entire chain from scratch. In Section~\ref{sec:runtime-model}, we show that this property also extends beyond static composition, where monotonicity carries over to runtime resource accesses, forming the basis for Phase~2 enforcement.


\section{Runtime Enforcement Model}
\label{sec:runtime-model}

Phase~1 composes tool policies as declared, but it operates only on the policies themselves, not on the values the agents will read, the sequence of calls it will make, or the accumulated exposure across these calls. This is because any given agent may choose resources dynamically based on its reasoning and input from other agents, both of which introduce irreducible variability into the workflow. For example, a coding agent might open whichever files match a search query that have been uploaded recently, or call whichever API endpoint another agent recommended, none of which can always (or almost ever) be known at session checkout. 

To address this gap, we define a runtime enforcement model that powers Phase~2, monitoring what the agent actually touches and intervening when the accumulated data sensitivity would make a subsequent tool call a policy violation. For example, a file reader authorized for \textsc{Confidential} data might only open \textsc{Public} documents in one session, or it might open a \textsc{Restricted} document that exceeds even its own declared level. Phase~1 cannot distinguish these cases, but Phase~2 can, because it resolves each resource's classification at access time. The model has four components: (i) a resource classification scheme that labels data at access time, (ii) a session taint state that accumulates sensitivity monotonically, (iii) a set of enforcement checks that trigger revocation when a policy violation is imminent, and (iv) a monotonicity extension that formally connects these runtime guarantees to the static invariant from \S\ref{sec:monotonicity}.

\paragraph{Resource classification.} The first component of Phase~2 assigns a sensitivity label to each piece of data the agent touches. Every resource in the system then carries a classification label drawn from the same ordered set used for tool policies ($\textsc{Public} < \textsc{Internal} < \textsc{Confidential} < \textsc{Restricted}$), along with an optional transmission prohibition flag. This then results in a set-up where when an agent invokes a tool on a specific resource, the system resolves that resource's label from a classification catalog at access time (though note that the resolved classification may exceed the tool's declared level, which would evade Phase~1's set-up). 

\paragraph{Session taint state.} The second component of Phase~2 tracks how sensitive the session has become as the agent acts over time. To do this, the session maintains a taint record $S = (C_s, \phi_s)$: a high-water classification $C_s$ that can only increase, and a sticky prohibition flag $\phi_s$ that, once set, cannot be cleared. On each resource access with resolved classification $C_r$ and prohibition flag $\phi_r$, the taint updates as:
\begin{equation}
\label{eq:taint-update}
  C_s \gets \max(C_s, C_r), \qquad \phi_s \gets \max(\phi_s, \phi_r)
\end{equation}
These are the same monotonic operators used in static composition (Equations~\ref{eq:classification} and~\ref{eq:prohibition}), now applied to the actual resources the agent touches rather than to policy declarations. As before, once elevated, neither field can be lowered for the remainder of the session.

\paragraph{Enforcement checks.} The next step outlines what to do with the accumulated exposure. Before each tool execution, the system evaluates three guards, any one of which can revoke the session. All three target the same class of violation (i.e., data crossing the organizational boundary when policy forbids it) but they detect it from different vantage points: First, if the session's taint record carries a transmission prohibition and the requested tool is outbound, second, if the resource being accessed in the current call itself prohibits external transmission and the chain includes outbound tools, or third, if the composed data flow from Phase~1 already prohibits external transmission and the current tool is outbound. When any of these guards fire, the system creates an attributed detection record identifying the violated NIST control, revokes the session immediately, and permits no further tool executions.

\paragraph{Monotonicity extension.} The fourth component ties these runtime mechanisms back to the formal guarantee from Phase~1: Because the taint update operators in Equation~\ref{eq:taint-update} mirror the composition operators in Invariant~\ref{inv:monotonicity}, the runtime model inherits the same structural property, where each resource access can only tighten the session's constraints but can never relax them. Together with Phase~1, this creates a defense-in-depth architecture in which the security boundary can only become more restrictive over the lifetime of a session.

Section~\ref{sec:implementation} describes the reference implementation that realizes both phases and characterizes their behavior on a representative tool catalog.


\section{Reference Implementation and Characterization}
\label{sec:implementation}

This section describes a reference implementation that realizes both phases of the architecture from Figure~\ref{fig:architecture} and characterizes their behavior on a representative tool catalog. We first describe the system's implementation (\S\ref{sec:system-overview}), then specify the tool catalog and enumeration methodology (\S\ref{sec:experimental-setup}), report Phase~1's static composition results (\S\ref{sec:results}), and demonstrate Phase~2's runtime enforcement on three illustrative scenarios (Figure~\ref{fig:playbooks}, \S\ref{sec:runtime-enforcement}).


\subsection{System Overview}
\label{sec:system-overview}

We built a reference implementation in C\# that integrates both phases into a single system. Phase~1 implements the MRS composition algorithm from \S\ref{sec:composition}, either issuing a time-limited session token or rejecting the request with an attributed explanation. Phase~2 implements the enforcement model from \S\ref{sec:runtime-model}, classifying resources at access time, updating the session's taint state, and revoking the session if any of the three guards fire. The policy store draws on 11 NIST SP~800-53 controls spanning five control families (AC, SC, AU, SI, MP), giving the composition engine a shared vocabulary of enforceable constraints across the catalog. 


\subsection{Experimental Setup}
\label{sec:experimental-setup}

To characterize Phase~1's static composition, we enumerate all policy combinations over a realistic tool catalog and measure how often MRS blocks them and why. Note that Phase~2's runtime enforcement is demonstrated separately in \S\ref{sec:runtime-enforcement}. We first describe the tool catalog, then outline the distribution of policies across classification levels, and finally describe the enumeration methodology.

\paragraph{Tool catalog.} Our test environment contains 32 tools drawn from two distinct categories. Ten \emph{infrastructure tools} are used to model common organizational capabilities (file readers, database query engines, HTTP clients, messaging endpoints, a code sandbox, a VPN gateway, a wiki reader, a cloud upload service, and an air-gapped lab), while twenty-two \emph{agent tools} mirror the tool set of a production coding agent (file system operations, shell execution, web access, task orchestration, planning, and scheduling). Each tool is bound to exactly one of 16 security policies, though multiple tools may share a policy when their security posture is identical.

\paragraph{Policy distribution.} The 16 policies span all four classification levels: Six operate at \textsc{Confidential} with internal-only flow (file reading, database access, code execution, VPN, file writing, agent orchestration), four are classified as \textsc{Internal} with internal-only flow (wiki reading, planning, worktree management, scheduling), three handle \textsc{Public} data with outbound flow (HTTP, Slack, cloud upload), and two govern \textsc{Restricted} data with internal-only flow and explicit transmission prohibition (bash execution, air-gapped lab). One policy, email, is classified as \textsc{Internal} but with outbound flow, making it the only policy that combines an internal classification with external-facing capability.

\paragraph{Enumeration.} Lastly, because MRS operates on policies rather than individual tools, we report results primarily at the policy level. The 16 policies yield $\binom{16}{2} = 120$ unordered pairwise combinations and $\binom{16}{3} = 560$ three-way combinations. For completeness, we also compute all $32 \times 31 = 992$ directed tool-level pairs and all $\binom{32}{3} = 4{,}960$ tool-level triples. Each combination is evaluated using the same composition logic described in \S\ref{sec:composition}.


\subsection{Results}
\label{sec:results}

\begin{figure*}[!t]
\centering
\includegraphics[width=\textwidth]{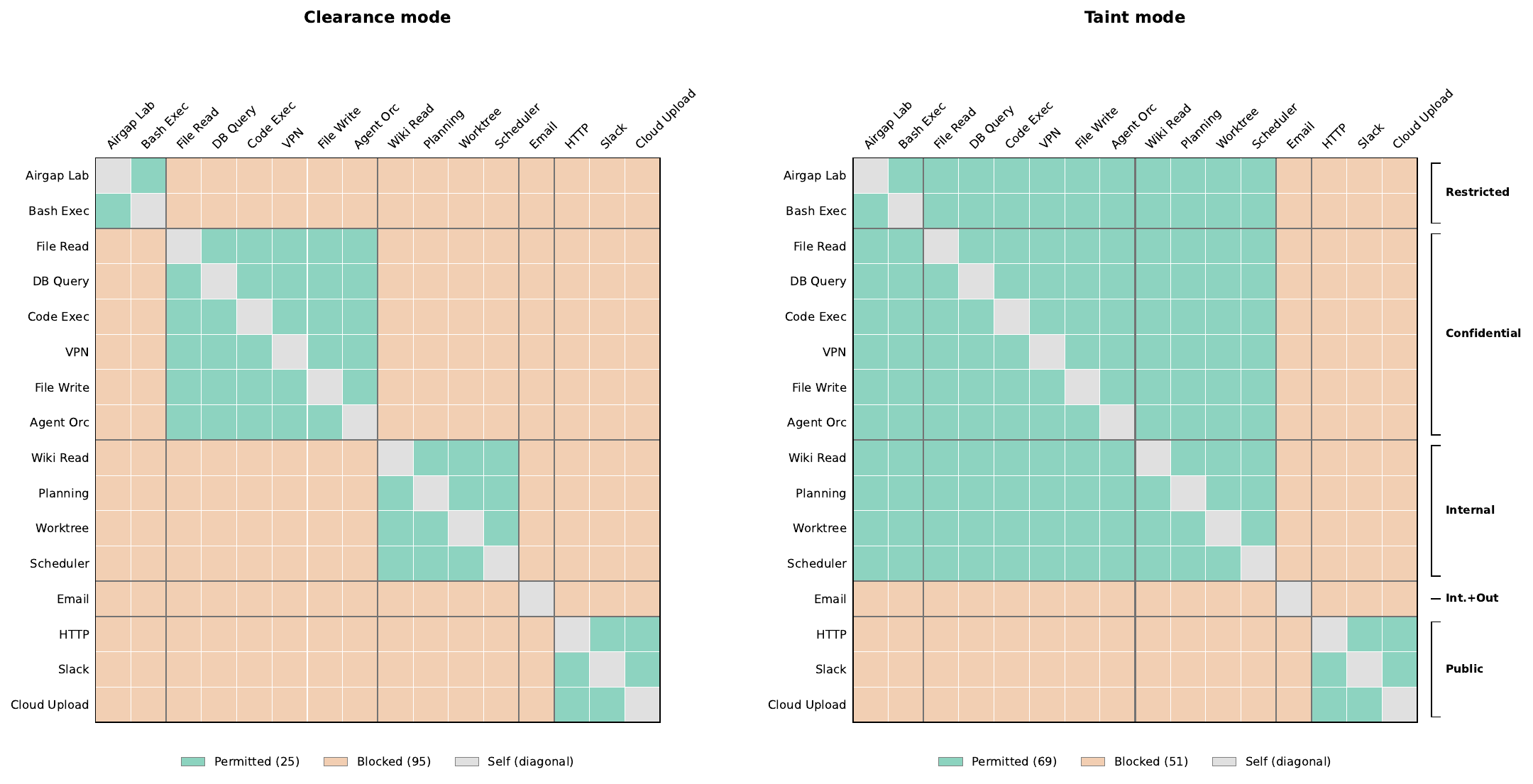}
\caption{Phase~1 composition under both modes (clearance and taint). \textit{Left:} Clearance mode (default): permitted pairs partition strictly into the four classification-level clusters along the diagonal. Green cells are permitted compositions, red cells are blocked, gray diagonal cells are self-pairs. \textit{Right:} Taint mode: the upper three classification levels (\textsc{Restricted}, \textsc{Confidential}, \textsc{Internal}) collapse into a single permitted region as the clearance check is gated off and mixed-classification chains execute with session taint elevated to the chain's high-water mark.}
\label{fig:heatmap}
\end{figure*}

Phase~1 blocks the majority of multi-policy compositions at session checkout, with the block rate depending on which composition mode (\S\ref{sec:composition}) is active. Clearance mode is the default and applies in regulated deployments where every tool must be cleared to the chain's high-water mark before participating, while taint mode is the alternative posture for mixed-classification pipelines where individual tools are not independently credentialed for the chain's peak sensitivity but organizational controls govern the pipeline as a whole. Table~\ref{tab:results} reports both modes across all pairwise and three-way combinations of the 16~policies and 32~tools in our catalog. Under clearance, MRS blocks 79.2\% of policy pairs and 95.5\% of triples; under taint, the same combinations block at 42.5\% and 60.5\% respectively (see Figure~\ref{fig:heatmap}). The block rate is higher at chain length~3 than at chain length~2 in both modes, consistent with the monotonicity invariant of \S\ref{sec:monotonicity}. The lower tool-level rate relative to policy-level reflects that many tools share policies and same-policy pairs are always permitted. 

\begin{table*}[!t]
\centering
\caption{Phase~1 composition results under both modes (clearance default, taint alternative) across all policy and tool combinations.}
\label{tab:results}
\small
\begin{tabular}{@{}lrrrrr@{}}
\toprule
 & & \multicolumn{2}{c}{\textbf{Clearance (default)}} & \multicolumn{2}{c}{\textbf{Taint}} \\
\cmidrule(lr){3-4} \cmidrule(lr){5-6}
\textbf{Combination type} & \textbf{Total} & \textbf{Blocked} & \textbf{Rate} & \textbf{Blocked} & \textbf{Rate} \\
\midrule
Policy pairs ($n = 2$)    & 120     & 95      & 79.2\% & 51      & 42.5\% \\[3pt]
Policy triples ($n = 3$)  & 560     & 535     & 95.5\% & 339     & 60.5\% \\[3pt]
Tool pairs (directed)     & 992     & 704     & 71.0\% & 322     & 32.5\% \\[3pt]
Tool triples (undirected) & 4{,}960 & 4{,}499 & 90.7\% & 2{,}350 & 47.4\% \\
\bottomrule
\end{tabular}

\smallskip
\begin{minipage}{0.95\textwidth}
\footnotesize\textit{Note.} Policy-level counts use unordered combinations ($\binom{16}{2}$ and $\binom{16}{3}$). Tool-level pairs are enumerated directionally ($32 \times 31$), though the current composition operators are all commutative, so each undirected pair produces the same verdict in both directions.
\end{minipage}
\end{table*}

\paragraph{Rejection mechanisms.} Under clearance mode, two mechanisms are responsible for all 95 blocked pairs. The clearance check in Step~3 is responsible for 91 (95.8\%): whenever two policies operate at different classification levels, the lower-classified policy is not authorized to handle data at the chain's high-water mark, and the composition is rejected. The remaining 4 pairs (4.2\%) are blocked by DENY-level boundary enforcement in Step~4, all involving the email sender paired with an \textsc{Internal}/internal-only policy. Here, both policies share a classification level, so the clearance check passes, but the email policy introduces SC-7 (Boundary Protection) at \textsc{Deny} while making the chain external-facing, triggering a boundary violation. The email policy is therefore incompatible with every other policy in the catalog.

Under taint mode, the clearance check is gated off (\S\ref{sec:composition}), so the 51 blocked policy pairs are handled by the downstream data-flow and DENY-enforcement checks. In clearance mode, it eats almost all classification-boundary cases before downstream checks see them, while in taint mode those checks see the full rejection workload.

\paragraph{Cluster structure.} Under clearance mode, the 25 permitted pairs partition into four classification-level clusters with no cross-cluster edges (Figure~\ref{fig:heatmap}, left panel): \textsc{Confidential} (6~policies, $\binom{6}{2} = 15$ pairs), \textsc{Internal} (4~policies, 6 pairs), \textsc{Public} (3~policies, 3 pairs), and \textsc{Restricted} (2~policies, 1 pair). The same pattern holds at chain length~3, where all 25 permitted triples are drawn from a single cluster ($20 + 4 + 1 + 0$). This structure follows directly from the clearance check: any pair spanning two classification levels is rejected, so composition is permitted only when all policies share the same level. The observed block rates are therefore a function of how policies are distributed across levels in this catalog, not an intrinsic property of the algorithm. Thus, what MRS guarantees structurally in clearance mode is that no cross-level composition is ever permitted, regardless of the distribution.

Under taint mode, the upper three classes (\textsc{Restricted}, \textsc{Confidential}, \textsc{Internal}) collapse into a single permitted region of 66 pairs (Figure~\ref{fig:heatmap}, right panel): 22 within-cluster pairs carry over from clearance ($15 + 6 + 1$) and 44 newly-permitted pairs span classes (12 \textsc{Restricted}$\times$\textsc{Confidential}, 8 \textsc{Restricted}$\times$\textsc{Internal}-non-email, 24 \textsc{Confidential}$\times$\textsc{Internal}-non-email). The remaining 3 permitted pairs lie inside the \textsc{Public} cluster, unchanged from clearance.  What survives in both modes is therefore exfiltration containment: data tainted above \textsc{Public} cannot reach an outbound tool, whether or not the clearance check is enforced.

\subsection{Runtime Enforcement}
\label{sec:runtime-enforcement}

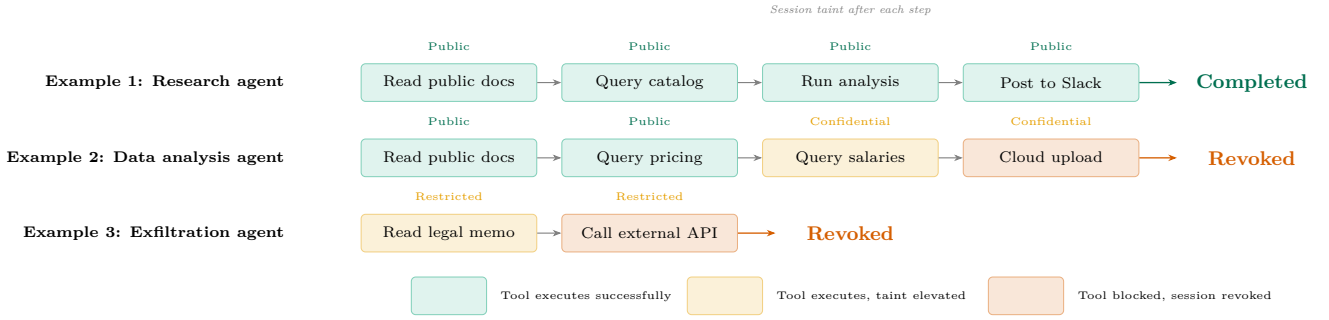
\begin{figure*}[!t]
\centering
\resizebox{\textwidth}{!}{%
\begin{tikzpicture}[
  stepbox/.style={draw, rounded corners=2pt, minimum height=0.6cm, font=\scriptsize, align=center, minimum width=2.8cm},
  ok/.style={stepbox, fill=oiGreen!12, draw=oiGreen!50},
  taint/.style={stepbox, fill=oiOrange!15, draw=oiOrange!50},
  kill/.style={stepbox, fill=oiVermillion!15, draw=oiVermillion!50},
  okout/.style={font=\small\bfseries, text=oiGreen!70!black},
  killout/.style={font=\small\bfseries, text=oiVermillion},
  label/.style={font=\scriptsize\bfseries, anchor=east},
  arr/.style={-{Stealth[length=4pt]}, line width=0.5pt, gray},
  taintlabel/.style={font=\tiny, anchor=south}
]
  \node[font=\tiny\itshape, gray, anchor=south] at (8.4, 3.35) {Session taint after each step};

  \node[label] at (-0.5, 2.4) {Example 1: Research agent};
  \node[ok] (r1) at (2.0, 2.4) {Read public docs};
  \node[ok] (r2) at (5.2, 2.4) {Query catalog};
  \node[ok] (r3) at (8.4, 2.4) {Run analysis};
  \node[ok] (r4) at (11.6, 2.4) {Post to Slack};
  \draw[arr] (r1) -- (r2);
  \draw[arr] (r2) -- (r3);
  \draw[arr] (r3) -- (r4);
  \node[taintlabel, text=oiGreen!70!black] at (2.0, 2.8) {Public};
  \node[taintlabel, text=oiGreen!70!black] at (5.2, 2.8) {Public};
  \node[taintlabel, text=oiGreen!70!black] at (8.4, 2.8) {Public};
  \node[taintlabel, text=oiGreen!70!black] at (11.6, 2.8) {Public};
  \draw[-{Stealth[length=4pt]}, oiGreen!70!black, line width=0.6pt] (r4.east) -- ++(0.6, 0);
  \node[okout] at (14.8, 2.4) {Completed};

  \node[label] at (-0.5, 1.2) {Example 2: Data analysis agent};
  \node[ok] (d1) at (2.0, 1.2) {Read public docs};
  \node[ok] (d2) at (5.2, 1.2) {Query pricing};
  \node[taint] (d3) at (8.4, 1.2) {Query salaries};
  \node[kill] (d4) at (11.6, 1.2) {Cloud upload};
  \draw[arr] (d1) -- (d2);
  \draw[arr] (d2) -- (d3);
  \draw[arr] (d3) -- (d4);
  \node[taintlabel, text=oiGreen!70!black] at (2.0, 1.6) {Public};
  \node[taintlabel, text=oiGreen!70!black] at (5.2, 1.6) {Public};
  \node[taintlabel, text=oiOrange] at (8.4, 1.6) {Confidential};
  \node[taintlabel, text=oiOrange] at (11.6, 1.6) {Confidential};
  \draw[-{Stealth[length=4pt]}, oiVermillion, line width=0.6pt] (d4.east) -- ++(0.6, 0);
  \node[killout] at (14.8, 1.2) {Revoked};

  \node[label] at (-0.5, 0.0) {Example 3: Exfiltration agent};
  \node[taint] (e1) at (2.0, 0.0) {Read legal memo};
  \node[kill] (e2) at (5.2, 0.0) {Call external API};
  \draw[arr] (e1) -- (e2);
  \node[taintlabel, text=oiOrange] at (2.0, 0.4) {Restricted};
  \node[taintlabel, text=oiOrange] at (5.2, 0.4) {Restricted};
  \draw[-{Stealth[length=4pt]}, oiVermillion, line width=0.6pt] (e2.east) -- ++(0.6, 0);
  \node[killout] at (8.4, 0.0) {Revoked};

  \node[ok, minimum width=1.2cm] (leg1) at (2.0, -1.0) {};
  \node[font=\tiny, anchor=west] at (2.7, -1.0) {Tool executes successfully};
  \node[taint, minimum width=1.2cm] (leg2) at (6.4, -1.0) {};
  \node[font=\tiny, anchor=west] at (7.1, -1.0) {Tool executes, taint elevated};
  \node[kill, minimum width=1.2cm] (leg3) at (11.2, -1.0) {};
  \node[font=\tiny, anchor=west] at (11.9, -1.0) {Tool blocked, session revoked};

\end{tikzpicture}%
}
\caption{Phase~2 (runtime enforcement): These three illustrative scenarios show how the session taint state evolves during tool execution across different plausible agentic deployment scenarios. Each box represents a tool invocation, with the session's accumulated classification level shown above in matching colour. Green boxes execute normally, orange boxes succeed but elevate the session taint to a higher classification, while red boxes are blocked because the accumulated taint makes the requested operation a policy violation. In Example~1, all resources are \textsc{Public}, so the session completes. In Example~2, querying employee salary data elevates the taint to \textsc{Confidential} with a transmission prohibition, so the subsequent cloud upload is blocked. In Example~3, reading a restricted legal memo immediately elevates the taint to \textsc{Restricted}, and the next outbound call is blocked.}
\label{fig:playbooks}
\end{figure*}

We demonstrate Phase~2 by walking through three scenarios (Figure~\ref{fig:playbooks}) that exercise the runtime enforcement model defined in \S\ref{sec:runtime-model}. Unlike Phase~1, which straightforwardly allows exhaustive enumeration over a finite set of policy combinations, Phase~2 depends on which resources an agent accesses at runtime, an input space that is almost unbounded and determined by the agent's reasoning as well as the input from sources the live internet. We therefore evaluate Phase~2 on a set of representative scenarios rather than exhaustive enumeration, selecting cases that primarily exercise Guard~1 (session-taint $\times$ outbound) and cover the spectrum from benign (no taint elevation) to immediate revocation. Each scenario below uses tools from the catalog in \S\ref{sec:experimental-setup} and shows how the taint state (Equation~\ref{eq:taint-update}) evolves with each resource access and when/why the enforcement checks fire.

\paragraph{Walkthrough.} In Example~1, a research agent reads public documents, queries a shareable product catalog, runs an analysis script, and posts a summary to Slack. As every resource it touches is classified as \textsc{Public}, the session taint stays at \textsc{Public} throughout and the agent completes its activities normally.

In Example~2, a data analysis agent starts the same way as the agent in Example~1 (reading public documents and querying product pricing). However, the agent then queries employee salary data, which is classified as \textsc{Confidential} with a transmission prohibition. The query itself succeeds because the database tool is cleared for \textsc{Confidential} data, but the session taint is now permanently elevated to \textsc{Confidential}. When the agent subsequently attempts a cloud upload, the system checks the accumulated taint against the tool. Because the cloud upload is an external-facing \textsc{Public} tool, but the session now carries \textsc{Confidential} data with a transmission prohibition, any outbound operation is a policy violation. Thus, the upload is blocked and the session is revoked.

In Example~3, an exfiltration agent reads a restricted legal memo on its very first tool call. The memo is classified as \textsc{Restricted} with a transmission prohibition, so the session taint is elevated to \textsc{Restricted} immediately. On the next call, the agent attempts to reach an external API. The taint check fires instantly because the session carries \textsc{Restricted} data under a transmission prohibition and the tool is external-facing. As such, the call is blocked and the session is revoked. The difference between this and Example~2 is that there is no window of normal operation, as the very first resource access sets the taint to a level that makes any subsequent outbound call impossible.

Overall, Phase~1 blocks incompatible tool chains before execution through static composition (\S\ref{sec:composition}), with the clearance mode producing the strict cluster partition reported above, and taint mode admitting mixed-classification chains while preserving the exfiltration boundary. Later, Phase~2 then catches violations that only become visible when the agent touches specific resources through runtime taint tracking (\S\ref{sec:runtime-model}). Across both modes, we have a monotonicity invariant of \S\ref{sec:monotonicity} as well as an exfiltration containment that keeps tainted data away from outbound tools, regardless of which mode is active. The mode choice is therefore a deployment-time decision about how to balance compositional reach against per-tool credentialing as opposed to a security-versus-no-security tradeoff. We discuss whether this balance strikes the right tradeoff between security and utility, along with the organizational and lifecycle considerations it raises for adopters, in §\ref{sec:discussion}.


\section{Discussion}
\label{sec:discussion}
The results in \S\ref{sec:results} show that DSCC produces a conservative security boundary by adopting an effective chain control set that is the most restrictive across all constituent tool policies. This introduces a tradeoff: DSCC limits the generality of an agent whenever its requested tool chain spans an incompatible policy requirement. We argue that this tradeoff is desirable, as the alternative default of no compositional enforcement introduces the multi-tool vulnerabilities surveyed before. Furthermore, decomposition into smaller tool chains abates issues with functionality limitations. Finally, the single effective control set of MRS gives auditors one policy artifact per session rather than a collection of tool-specific policies with no cross-tool enforcement.

The remainder of this section examines what follows from this design choice. First, we examine the utility-security tradeoff DSCC imposes, arguing that the cost of over-restriction is recoverable while the cost of under-restriction is not (§\ref{sec:utility-security}). Second, we chart how DSCC policies should be organizationally authored, owned, and maintained over time (§\ref{sec:policy-lifecycle}). Third, we turn to a specific advantage of DSCC in mitigating frontier risks of agentic systems, arguing that its fail-closed composition posture enables it to function as a control-level complement to emerging capability-level guardrails (§\ref{sec:frontier}). 


\subsection{On the Utility-Security Tradeoff}
\label{sec:utility-security}

As §\ref{sec:results} showed, for our reference implementation operating in clearance mode, the clearance check accounts for 95.8\% of MRS's rejections by blocking compositions that span two conflicting classification levels. Within a single classification cluster, nearly all pairs and triples are permitted (the four exceptions all involve the email policy, which carries SC-7 \textsc{Deny} regardless of cluster), while across clusters, none are. Under taint mode the same boundary is enforced downstream rather than upstream, with the rejection rate falling from 95.5\%/79.2\% to 60.5\%/42.5\% on triples/pairs (Table~\ref{tab:results}). The effective cost of DSCC to a deployment therefore depends on how its tool catalog distributes policies across classification levels relative to its intended workflows. A deployment whose policy authors have placed tools at appropriate levels sees a low block rate. Conversely, coarsely written policies or workflows that trigger cross-classification boundaries will see a higher block rate.

When MRS blocks a chain, two readings are possible. The first is that the block did precise work, preventing the chain from crossing a boundary that it should not. The second, unintended case is that MRS overblocked legitimate workflows, as the policy was too coarse. In the latter case, one or more of the tool policies need refining. MRS itself does not distinguish between these two cases. However, it reports to the user which control binding  caused the denial, allowing them to examine the policy in question and consider refinement, where appropriate.

We propose this tradeoff through MRS since the cost of these two errors is asymmetrical. A legitimate chain that gets blocked by an overly coarse policy disrupts user experience, but is monitorable and amendable. An unsafe chain that slips through an under-specified policy is different in that it cannot be reconciled post-hoc. DSCC prioritizes the recoverable cost over the unrecoverable one, only adding operational overhead in reviewing policies that likely would have already needed refinement. We return to this asymmetry in §\ref{sec:frontier}, where it underlies the case for enabling compositional enforcement by default.


\subsection{DSCC Policy Lifecycle}
\label{sec:policy-lifecycle}

DSCC requires a tool catalog to run, and each tool needs a classification level, a control binding to the framework of choice (NIST SP~800-53 in our implementation; see §\ref{sec:implementation}), and any data-flow constraints that apply (i.e. a use policy). Authoring DSCC policies is  a continuous task throughout the tool lifecycle, requiring  ongoing monitoring, policy reviews and refinement as tools and workflows evolve. Tools acquire new versions, based on different training data and introducing new capabilities over time. Their policies must change with them, in a manner that accounts for organizational knowledge and incentives.

The model assumes a federated division of labor. A central security team owns the shared vocabulary, which includes the
classification scheme, the control framework, and the classification floors. In parallel, tool developers author the policy for their respective tool. Two considerations  underpin the split. First, the developers who built a tool understand its data access and externalities better than  central reviewers, which puts them in the best position to
describe the tool's limitations. Second, developers are typically  incentivized towards the most permissive policy that allows their tool to run seamlessly, requiring that a separate objective function oversees the audit trail.  Federation thus balances developer-supplied accuracy with centrally-owned minimal requirements and audit.
 
This federative structure also shapes what happens when a chain is denied. Because MRS records which binding in which tool caused the denial, triage starts from a single tool rather than the whole chain. Usually the developer whose chain got blocked is the one who is most incentivized to resolve it, but the binding can only be relaxed by whoever owns the blocking policy. Federation does not resolve these cases, but it narrows the candidate set from an organization-wide search to the small number of parties whose tools or boundaries are impacted by the denial.


\subsection{DSCC as a Mitigation Lever for Frontier Risk}
\label{sec:frontier}

A specific emerging area where DSCC can serve as a mitigation layer is frontier AI risks, a class of severe, potentially large-scale harms associated with the most capable general-purpose models. These risks are increasingly the subject of a dedicated governance ecosystem spanning developer-led safety frameworks
\cite{microsoft2025fgf,deepmind2025fsf,anthropic2025rsp,openai2025preparedness}, emerging regulation and standards \cite{euaiact2024,nist2024misuse}, and independent AI safety reports \cite{bengio2026international}. A common
thread across these efforts is the identification of dangerous capabilities that require dedicated evaluation and mitigation once a model crosses a pre-defined risk threshold \cite{shevlane2023model,kinniment2024evaluating}. Autonomous, multi-step tool use is consistently named among these capabilities. This enables agents to chain actions in manners that are harmful, including chaining individually benign actions into outcomes a human principal would not have necessarily authorized. 

DSCC operates on a different axis than the evaluations that dominate frontier governance. Frontier risk governance capability evaluations monitor what a model can do, probing whether it is able to plan autonomously, chain tools, or evade oversight. Instead, DSCC instead constrains what the tool is permitted to do, irrespective of the model's capabilities, making these two axes orthogonal. A capability evaluation is incomplete as it samples from an unbounded behavior space and establishes only a lower bound on what the model can do for specific and anticipated risk areas. MRS composition, by contrast, confines the agent to the policy space
it governs. As the findings in \S\ref{sec:results} show, every admissible combination of tool policies has a fixed, pre-computable verdict. DSCC therefore does not attempt to predict the agent's behavior, but bounds it based on continuously refined action policies.

DSCC does not interfere with the intent or planning capabilities of an agent, but it limits the action space available to it. A chain that would cross a classification boundary is blocked whether the agent autonomously chose to cross it or was instructed into it, since the check only looks at the declared tool policies. In this sense, DSCC acts as a circuit breaker at the policy level, capping how far an agent's tool use can reach. It does so independently of the agent's reasoning, instilling a control-level backstop that sits alongside capability-level evaluations and mitigations, rather than duplicating them.

In the context of frontier risk, DSCC can be deployed in two ways, which differ in the extent and triggers for its enforcement. The first is \emph{always-on}: DSCC runs
on every agent session, regardless of the model capability. This is a safe default because, as \S\ref{sec:utility-security} argued, blocking a good chain is recoverable but letting a bad one through is not and because the results in
\S\ref{sec:results} show that chains within a single classification level are almost always permitted, the cost of running by default is low. The second is \emph{threshold-triggered}: DSCC is switched on, or made stricter, only once a capability evaluation shows the model has crossed a risk tier defined by a frontier framework (see, e.g.,~\cite{microsoft2025fgf}). This allows the deployer to tailor DSCC application to the level of risk. Where the risks are negligible for the implementation of DSCC, it supports different postures triggered by indications of higher-risk capabilities.


\clearpage
\bibliographystyle{unsrtnat}
\bibliography{references}

\clearpage
\appendix
\renewcommand{\thesection}{}
\renewcommand{\thesubsection}{}
\section*{Appendix}
\label{app:catalog}
\addcontentsline{toc}{section}{Appendix}

\captionsetup{singlelinecheck=false, justification=raggedright, hypcap=false}

This appendix provides the complete data underlying the evaluation in \S\ref{sec:implementation}. Table~\ref{tab:controls} lists the 11 NIST SP~800-53 controls used in the reference implementation, of which two (AC-4 and SC-7) default to \textsc{Deny} and the remaining nine to \textsc{Restrict}. Table~\ref{tab:policies} gives the full definitions of all 16 security policies, including classification level, flow direction, transmission prohibition, session TTL, and assigned controls. Table~\ref{tab:tools} maps each of the 32 tools to its governing policy, grouped by category. Together, these three tables are sufficient to reproduce the enumeration results reported in \S\ref{sec:results}.

\vspace{0.8em}

\begin{minipage}{\textwidth}
\captionof{table}{NIST SP 800-53 controls used in the reference implementation.}
\label{tab:controls}
\scriptsize
\begin{tabular}{@{}llll@{}}
\toprule
\textbf{Family} & \textbf{ID} & \textbf{Name} & \textbf{Default level} \\
\midrule
AC & AC-3 & Access Enforcement & Restrict \\
AC & AC-4 & Information Flow Enforcement & Deny \\
AC & AC-6 & Least Privilege & Restrict \\
SC & SC-7 & Boundary Protection & Deny \\
SC & SC-8 & Transmission Confidentiality \& Integrity & Restrict \\
SC & SC-13 & Cryptographic Protection & Restrict \\
SC & SC-28 & Protection of Information at Rest & Restrict \\
AU & AU-2 & Event Logging & Restrict \\
AU & AU-6 & Audit Record Review \& Reporting & Restrict \\
SI & SI-4 & System Monitoring & Restrict \\
MP & MP-5 & Media Transport & Restrict \\
\bottomrule
\end{tabular}
\end{minipage}

\vspace{1.2em}

\begin{minipage}{\textwidth}
\captionof{table}{Complete policy catalog used in the evaluation.}
\label{tab:policies}
\scriptsize
\begin{tabular}{@{}llllrl@{}}
\toprule
\textbf{Policy} & \textbf{Classification} & \textbf{Flow} & \textbf{Prohibit} & \textbf{TTL (h)} & \textbf{Controls} \\
\midrule
File Reader       & Confidential & Internal & No  & 48 & AC-3, AC-4 \\
Database Query    & Confidential & Internal & No  & 24 & AC-3, AC-6 \\
Code Interpreter  & Confidential & Internal & No  & 12 & AC-3, AC-6, SI-4 \\
VPN Gateway       & Confidential & Internal & Yes & 8  & AC-3, SC-7, SC-8, SC-13 \\
File Writer       & Confidential & Internal & No  & 24 & AC-3, AC-6, AU-2 \\
Agent Orchestrator & Confidential & Internal & No  & 24 & AC-3, AC-4, AU-2 \\
Wiki Reader       & Internal     & Internal & No  & 48 & AC-3, AC-4 \\
Planning          & Internal     & Internal & No  & 48 & AC-3 \\
Worktree          & Internal     & Internal & No  & 24 & AC-3, AC-6 \\
Scheduler         & Internal     & Internal & No  & 48 & AC-3, AU-2 \\
Email Sender      & Internal     & Outbound & No  & 48 & SC-7, SC-8 \\
HTTP Client       & Public       & Outbound & No  & 48 & SC-7, SC-8 \\
Slack Notifier    & Public       & Outbound & No  & 48 & SC-7 \\
Cloud Upload      & Public       & Outbound & No  & 4  & SC-7, SC-8, SC-13, MP-5 \\
Bash Executor     & Restricted   & Internal & Yes & 4  & AC-3, AC-6, AU-2, SI-4 \\
Air-gapped Lab    & Restricted   & Internal & Yes & 2  & AC-3, AC-4, AC-6, SC-28, SI-4 \\
\bottomrule
\end{tabular}
\end{minipage}

\vspace{1.2em}

\begin{minipage}{\textwidth}
\captionof{table}{Complete tool catalog used in the evaluation.}
\label{tab:tools}
\scriptsize
\begin{tabular}{@{}ll@{\hspace{3em}}ll@{}}
\toprule
\textbf{Tool} & \textbf{Policy} & \textbf{Tool} & \textbf{Policy} \\
\midrule
\multicolumn{4}{@{}l}{\emph{Infrastructure tools (10)}} \\
Read Documents        & File Reader       & Send Email            & Email Sender \\
Query Database        & Database Query    & VPN Access            & VPN Gateway \\
Web API Call          & HTTP Client       & Read Wiki Pages       & Wiki Reader \\
Send Slack Message    & Slack Notifier    & Cloud File Upload     & Cloud Upload \\
Run Code (Sandbox)    & Code Interpreter  & Detonation Environment & Air-gapped Lab \\[3pt]
\multicolumn{4}{@{}l}{\emph{Agent tools (22)}} \\
Agent                 & Agent Orchestrator & WebFetch             & HTTP Client \\
TaskOutput            & Agent Orchestrator & WebSearch            & HTTP Client \\
TaskStop              & Agent Orchestrator & TodoWrite            & Planning \\
Bash                  & Bash Executor      & AskUserQuestion      & Planning \\
Glob                  & File Reader        & Skill                & Planning \\
Grep                  & File Reader        & EnterPlanMode        & Planning \\
Read                  & File Reader        & ExitPlanMode         & Planning \\
Edit                  & File Writer        & EnterWorktree        & Worktree \\
Write                 & File Writer        & ExitWorktree         & Worktree \\
NotebookEdit          & File Writer        & CronCreate           & Scheduler \\
                      &                    & CronDelete           & Scheduler \\
                      &                    & CronList             & Scheduler \\
\bottomrule
\end{tabular}
\end{minipage}

\end{document}